\documentclass[12pt]{article}
\usepackage[dvips]{graphics}
\usepackage[english]{babel}
\usepackage{amssymb}
\usepackage{amsmath}
\usepackage{graphicx}
\usepackage[applemac]{inputenc}
\usepackage{cite}
\usepackage{subfigure}
\def\beq{\begin{equation}}
\def\eeq{\end{equation}}
\def\bea{\begin{eqnarray}}
\def\eea{\end{eqnarray}}
\def\nn{\nonumber}
\def\sN{\mathbf{\bar N}}
\def\sL{\mathbf{L}}
\def\sH{\mathbf{H_u}}
\def\nn{\nonumber}

\def\sneu{\tilde{\nu}_R}

\begin{document}

\title{Non-thermal right-handed sneutrino dark matter and the $\Omega_{DM}/\Omega_b$ problem}

\author{V\'eronique Pag\'e \\
    {\em Centre for Particle Theory,} \\
{\em University of Durham, Durham, DH1 3LE, UK}}

\maketitle

\abstract{We argue that the superpartner of the Dirac right-handed neutrino is a prime candidate for dark matter created from a 'mattergenesis' mechanism.  We show that due to the smallness of the Yukawa couplings, a right-handed sneutrino density created in the early Universe would not be erased by annihilations, which remain out of thermal equilibrium.  It would also not be drowned by a later, additional production of right-handed sneutrinos, as the relic density of the non-thermal right-handed sneutrinos is found to be generally negligible compared to the observed dark matter density.  Mild constraints on sneutrino masses and trilinear SUSY-breaking couplings are obtained.  Possible mattergenesis scenarios are also mentioned}


\section{Introduction}
While the energy density of baryonic matter, dark matter (DM) and dark energy 
are being measured more precisely, the very nature and origin of 
dark matter remain open questions.  In the popular neutralino dark matter scenario, 
the dark matter density is equated to the LSP (lightest supersymmetric partner of the MSSM) neutralino relic density  \cite{goldberg_relic,ellis_relic}.  Indeed, it is often considered to be one of the many appealing aspects of the MSSM that it naturally provides a dark matter candidate.  This scenario, however, fails to address the intriguing question of the ratio of the dark and baryonic matter energy densities.  We now know these to be given by \cite{wmap2006}
\beq \label{eq: omegas}
\frac{\Omega_{DM}}{\Omega_b}\sim \frac{0.19}{0.04} \sim 4.9 ~. 
\eeq
The baryon relic density is known to be much smaller than $4\%$ of the critical energy density (see eg. \cite{kolbturner}) and for this reason it is necessary to find a baryogenesis mechanism that can create the right amount of baryons.  In the common neutralino dark matter scenario, obviously, no such thing is necessary for DM since, as we said, the relic density of neutralino is directly the DM density.  Since the origins of both types of matter are unrelated, the similarity in their present-day densities appears simply as a coincidence.  This problem, sometimes called the $\Omega_{DM}/\Omega_b$ problem, was first mentioned by  \cite{dmbarr1}, and has received increasing attention  \cite{dmbarr2, dmkaplan, dmthomas, dmkuzmin,  dmkusenko,dmfarrar, dmkitano1, 
dmhooper, dmkitano2, tytgat, abelpage}.  The usual route to tackle this problem is to try and create a mechanism that generates both kinds of matter, baryonic and dark, at the same time.  

It has been suggested before (eg. in \cite{tytgat, dmkuzmin}) that candidates for mattergenesis-induced DM should generally have weak or even super-weak interactions with the 'visible' sector.  If the candidate never thermalises, the asymmetry created by mattergenesis will not be erased or reprocessed at later times.  In this case, the smallness of the couplings acts as a built-in protection of the DM asymmetry.  A condition that other possible sources of DM stay small is also added.  Another protection mechanism has been suggested in \cite{dmthomas}, where this time the candidate is thermal in the early Universe but freezes-out at some temperature $T\sim m_{DM}/20$, creating a low relic density.  The observed DM density (and baryon density) is created after freeze-out by the decay of a heavier particle which couples to both dark and baryonic matter.  For this reason the mechanism has been called the 'late decay' scenario.  In both cases, the DM asymmetry is created at a time when the DM candidate is out of thermal equilibrium with the plasma, and will remain so.  Common to both cases is also the fact that other sources of the DM candidate have to be kept small.

While there is no candidate in the Standard Model or even in the MSSM that has a low relic density and never thermalises, once these models are augmented by right-handed (s)neutrinos to explain the observed neutrino masses the situation is somewhat different.  We will argue here that the MSSM+Dirac right-handed (s)neutrino offers a prime DM candidate for mattergenesis scenarios in the sterile sneutrino LSP.   Due to the smallness of its Yukawa coupling, the right-handed (RH) sneutrino is completely decoupled from the thermal plasma and has typically a low relic density.  In the 'late decay' scenario, the condition that the heavy particle should decay after the candidate freezes-out would be lifted.  This implies that for \emph{any} mattergenesis scenario creating the Dirac RH sneutrino as DM, the DM asymmetry would neither be erased by fast annihilation processes nor be drowned by later freeze-out creation.

In the following we will first introduce a Dirac mass term for the (s)neutrino in the MSSM and extract from the model the RH sneutrino interactions.  In section \ref{sec: nontherm} we will show that under fairly mild constraints on left- and right-handed sneutrino mixing the RH sneutrino remains out of equilibrium.  We will obtain the RH sneutrino relic density numerically in section \ref{sec: relicdensity} and show that it is much lower than the observed DM density for natural choices of parameters.  Throughout we shall suppose that a hypothetical mattergenesis scenario has created RH sneutrino DM and baryons in quantities mentioned in eq.(\ref{eq: omegas}).  Such mattergenesis scenarios already exist in the literature: an Affleck-Dine mattergenesis scenario with RH sneutrino DM was proposed in \cite{abelpage} and in \cite{mcdonald}, and earlier an 'early-decay' mattergenesis scenario was suggested in \cite{dmkuzmin}, where it was noticed that the simplest implementation might be in the MSSM+$\sneu$.  We will show in the last section that indeed the parameters these scenarios necessitate do imply a non-thermal right-handed sneutrino with low relic density.

\section{RH sneutrino interactions} \label{sec: model}
We add to the MSSM model a RH neutrino superfield $\mathbf{\bar N}$ which is given a Dirac mass term in the superpotential $\mathcal{W}$:
\beq \label{superpotential}
\mathbf{\mathcal{W}} \supset  \lambda \sL^i \epsilon_{ij} \sH^j \sN ~,
\eeq
where $\mathbf{L}$ is the left-handed (LH) lepton doublet superfield and $\mathbf{H_u}$ is the up-type Higgs superfield.  We also now have new SUSY-breaking terms
\bea
\mathcal{V_{SB}}= m_{\tilde{\nu}_L}^2 \tilde{\nu_L}^*\tilde{\nu_L} + m_{\sneu}^2\tilde{\nu_R}^{c*}\tilde{\nu_R}^c + \left(a \lambda h_u \tilde{l} \sneu^c +h.c. \right)
\eea
where $a$ is a mass dimension trilinear coupling, and $h_u$ ($l$) is the up-type higgs (lepton) $SU(2)$ doublet.  The RH sneutrino has been introduced as a gauge singlet and as such will only have a handful of interactions.  $F$-terms are the source of 4-point interactions and higgsino exchange (keeping only terms involving the RH sneutrino):
\bea
\mathcal{L}_F &=& \sum_{i=L,H_u,N_R}\left| \frac{\partial W}{\partial \Phi_i} \right|^2+\sum_{i,j=L,H_u,N_R}\frac{\partial^2 W}{\partial \Phi_i \Phi_j}  \psi_i\psi_j \nn \\
&=& -\lambda^2 \left(\tilde{e}\tilde{e}^*\sneu^c \sneu^{c*}+h_u^+h_u^{+*}\sneu^c\sneu^{c*} +\tilde{\nu_L}\tilde{\nu_L}^*\sneu^c\sneu^{c*}+h_u^0h_u^{0*}\sneu^c\sneu^{c*} \right) \nn \\
& & + \lambda \left( \sneu^c\tilde{H_u}^+e - \sneu^c\tilde{H_u}^0\nu_L  \right) ~,
\eea
where all $\Phi$'s (resp. $\psi$'s) stand for the scalar (resp. fermionic) part of the superfields, and $h^{+,0}_u$ is the higgs boson, while $\tilde{H}^{+,0}_u$ is the higgsino.   The trilinear term in the SUSY-breaking Lagrangian provides higgs interactions:
\bea
\mathcal{L}_{SB,\sneu} = a \lambda \left( h_u^+\tilde{e}\sneu^c - h_u^0\tilde{\nu_L}\sneu^c  +h.c. \right) ~.
\eea
These are all the interactions the \emph{sterile} sneutrino has.  However once the electroweak symmetry is broken, the higgs acquires a \emph{vev}, and the left- and right-handed sneutrino mix \footnote{It could also be said that before the electroweak phase transition, left- and right-handed sneutrinos mix via 4-point interactions, higgsino exchange or higgs exchange.  To call these a 'mixing' phenomenon or simply interactions is in fact a matter of timescale; would they  be fast enough, one could say that the RH sneutrino effectively has a left-handed part. We will see however in the next section that left-right mixing is in fact very slow compared to the Universe expansion before the electroweak phase transition, and indeed never equilibrate.  This need not be the case once the higgs has acquired a \emph{vev}; hence our choice of wording.}.  Indeed, considering this time only mass terms in the Lagrangian:
\bea
\mathcal{L}_{mass}= - m_{\sneu}^2\tilde{\nu_R}^{c*}\tilde{\nu_R}^c - m^2_{\tilde{\nu_L}}\tilde{\nu_L}^*\tilde{\nu_L} - a \lambda v \sin \beta \tilde{\nu_L}\sneu^c - a^* \lambda v \sin \beta \tilde{\nu_L^*}\sneu^{c*}
\eea
where the up-type higgs has been replaced by its \emph{vev},
\bea
\left(\left< h_u^+\right>,\left< h_u^0\right>  \right)=\left(0,v\sin\beta \right)~,
\eea
the usual procedure leads to the following mass-eigenstate RH sneutrino:
\bea \label{eq: rm1}
\tilde{\nu}_{RM}=\frac{1}{\sqrt{\left( a\lambda v\sin\beta \right)^2+\left( m^2_{\tilde{\nu_L}}-m_{\sneu}^2 \right)^2}}\left[ \left(  m^2_{\tilde{\nu_L}}-m_{\sneu}^2\right)\sneu -a \lambda v\sin\beta \tilde{\nu_L} \right]~.
\eea
For simplicity reasons, we will take $\sin \beta =1$ from now on.  As can be expected, we will show in the next section that for the RH sneutrino never to thermalise, it is necessary that the LH (active) part of it be small compared to the sterile part ; in this case eq.(\ref{eq: rm1}) can be rewritten
\bea \label{eq: LRsneu}
\tilde{\nu}_{RM} \simeq \sneu - \frac{\lambda a v}{\left( m^2_{\tilde{\nu_L}}-m_{\sneu}^2 \right)} \tilde{\nu_L} ~.
\eea

\begin{figure}[t]
\begin{center}
\begin{tabular}{c c} 
\begin{minipage} {0.45\textwidth}   \includegraphics[width=0.9\textwidth]{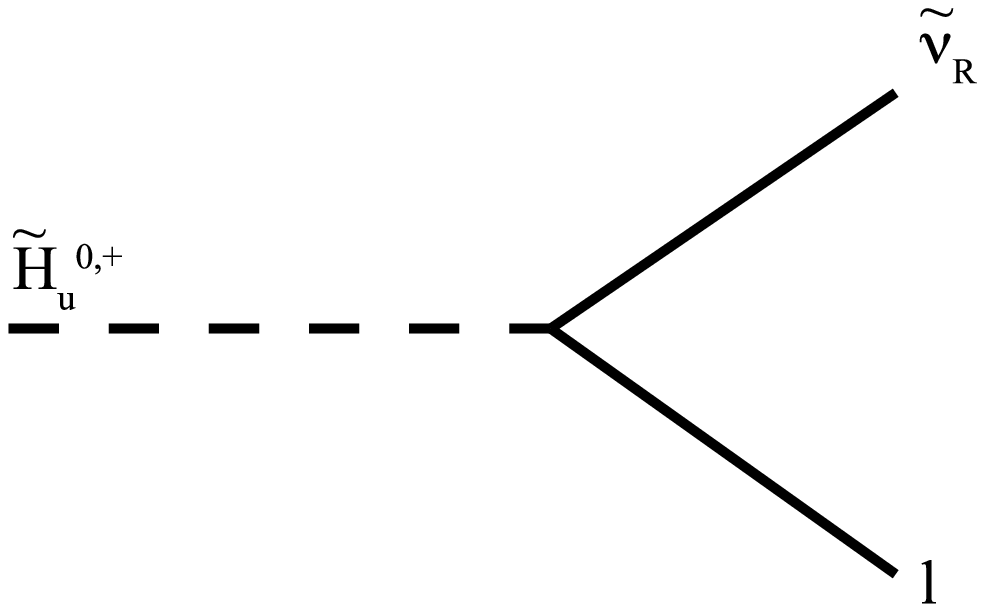}  \end{minipage}
& \begin{minipage} {0.45\textwidth} \includegraphics[width=0.9\textwidth]{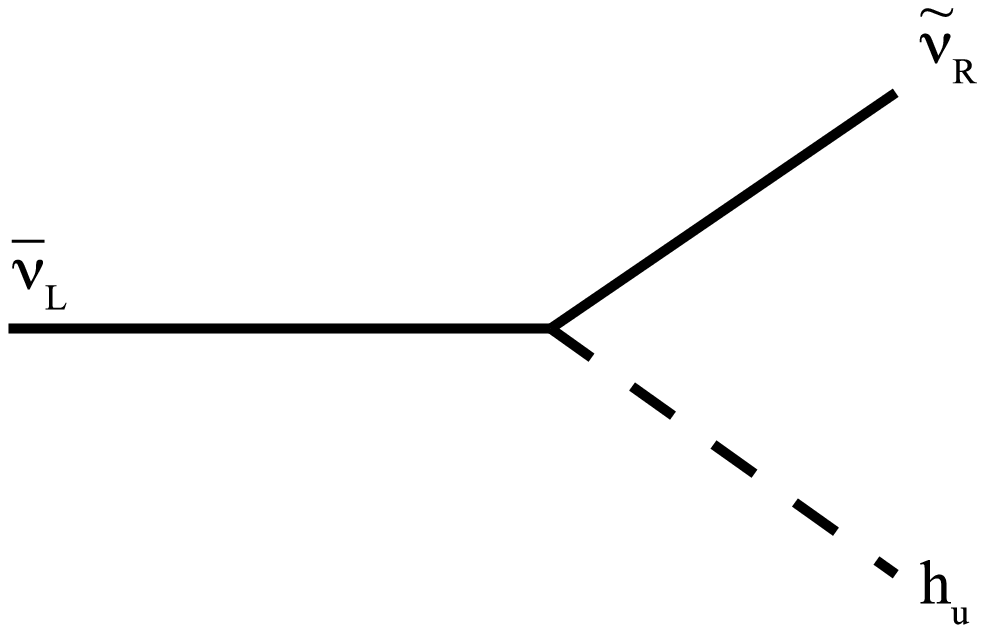}\end{minipage} \\
 $\left|M\right|^2 = 4 \lambda^2 \left( p_Hp_{l}- m_Hm_{l}  \right)$ & $\left|M\right|^2 = \lambda^2 a^2$\\
\begin{minipage}{0.45\textwidth} \includegraphics[width=0.9\textwidth]{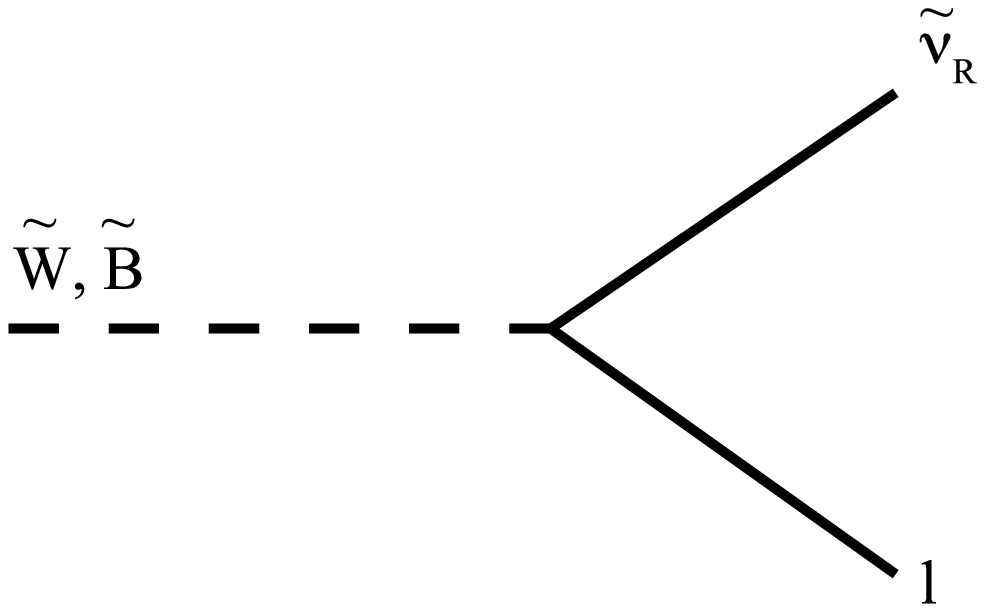} \end{minipage} 
& \begin{minipage}{0.45\textwidth} \includegraphics[width=0.9\textwidth]{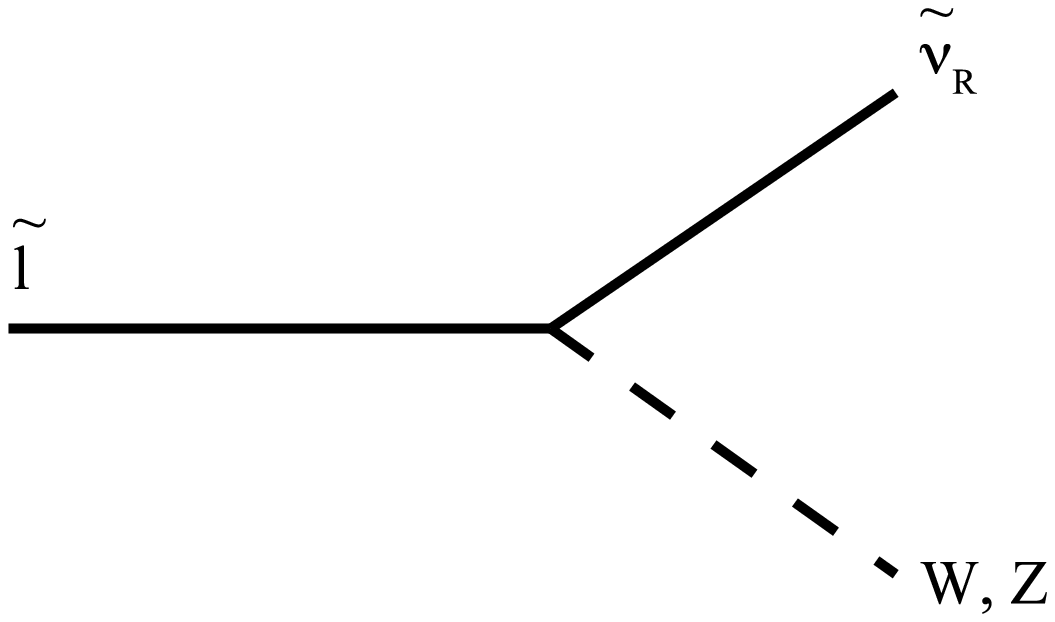}\end{minipage} \\ 
 $\left|M\right|^2 = 8 \lambda^2 v^2 A^2 \left( p_{\tilde{W},\tilde{B}}\cdot p_{l}- m_{\tilde{W},\tilde{B}}m_{\nu_l}  \right)$ &  $\left|M\right|^2 = \lambda^2 M_{W,Z}^2 A^2$ {\Large[} $- \left( p_{\tilde{\nu_R}}+ p_{\tilde{l}}  \right) \cdot \left( p_{\tilde{\nu_R}}+ p_{\tilde{l}}  \right)$\\
&$+\frac{1}{M^2_{W,Z}}  \left(p_{\tilde{\nu_R}}+p_{\tilde{l}}  \right) \cdot p_{W,Z}  \left(p_{\tilde{\nu_R}}+p_{\tilde{l}}  \right) \cdot p_{W,Z}$ {\Large]} \\
\end{tabular}
\caption{RH sneutrino interactions with their corresponding amplitude, with $A^2$ defined as $A^2\equiv a^2/\left( m_{\tilde{\nu_L}}^2- m_{\tilde{\nu_R}}^2\right)^2$.  4-points interactions amplitudes are all simply proportional to $\lambda^4$}
\label{fig: interactions}
\end{center}
\end{figure}

In the following we will drop the subscript $M$ when speaking of $\tilde{\nu}_{RM}$, and simply consider $\tilde{\nu}_R$ to have a small left-handed component when left-right equilibration processes are in equilibrium.  Thus the RH sneutrino inherits the gauge and gaugino interactions of the LH sneutrino (see for example \cite{martinprimer}):
\bea
\mathcal{L}_{gaugino-\sneu} = & &- \sqrt{2} g_2\frac{\lambda a v}{\left( m_{\tilde{\nu}_L}^2-m_{\sneu}^2 \right)}  \left[  \sneu^c \nu_L \tilde{W}^0 + \sneu^{c*} e \tilde{W}^+ + h.c.\right] \nn \\
& & - \sqrt{2}g_1\frac{\lambda a v}{\left( m_{\tilde{\nu}_L}^2-m_{\sneu}^2 \right)} \left[  \sneu^{c*} \nu_L \tilde{B} + h.c. \right] ~.
\eea
\bea
\mathcal{L}_{gauge-\sneu}&=& -i  \lambda \frac{\sqrt{2} M_W a}{\left( m_{\tilde{\nu}_L}^2-m_{\sneu}^2 \right)} \left( W^+ \sneu^{c*} \overleftrightarrow{\partial} \tilde{e} + W^- \tilde{e}^* \overleftrightarrow{\partial} \sneu^c \right) \nn \\
& & -i \lambda \frac{M_Z a}{\left( m_{\tilde{\nu}_L}^2-m_{\sneu}^2 \right)} Z \tilde{\nu}_L^{*} \overleftrightarrow{\partial} \sneu^c ~.
\eea
The list of the RH sneutrino interactions (excluding 4-points interactions) is included in fig.(\ref{fig: interactions}).  It is interesting to note that in the MSSM + Dirac (s)neutrino, while the RH neutrino remains completely sterile, the RH sneutrino can mix with its LH counterpart and through it interact with the gauge sector.

Before continuing we should mention what happens when one introduces a Majorana mass for the RH (s)neutrino.  The superpotential is now
\bea
\mathbf{\mathcal{W}} \supset  \lambda \sL^i \epsilon_{ij} \sH^j \sN + M_R \sN \sN
\eea
which results in the following added $F$-terms in the Lagrangian:
\bea
\mathcal{L}_{F,majorana}=-2M_R^2 \sneu^c\sneu^{c*}+\left( M_R\lambda \tilde{e}h^+\sneu^{c*} -M_R\lambda \tilde{\nu}_Lh^0\sneu^{c*}+h.c. \right) ~.
\eea
Adding the SUSY-breaking terms,
\bea
\mathcal{L}_{majorana}=&-&\left( m_{\tilde{\nu}_R}^2+2M_R^2 \right) \sneu^c\sneu^{c*}  - m_{\tilde{\nu}_L}^2 \tilde{\nu}_L^*\tilde{\nu}_L \nn\\
&+& \lambda \left( M_R+ a \right)\left( \tilde{e}h^+\sneu^{c*} - \tilde{\nu}_Lh^0\sneu^{c*}  +h.c. \right)~.
\eea
We want only to consider here the case of the RH sneutrino as the LSP; then automatically this excludes a very large, see-saw type of $M_R$.  A Majorana mass small enough for the RH sneutrino to be the LSP and that does not allow for thermalisation might be possible, as one can infer from the next section.  Indeed it was shown recently that the right-handed sneutrinos can be non-thermal in the presence of a Majorana mass \cite{degouvea}.   Such cases of pseudo-Dirac neutrinos are highly constrained from a phenomenological point of view (see for example \cite{abazajian}).  We shall not consider this case any further.

\section{(Non-) thermalisation of the RH sneutrino} \label{sec: nontherm}
Let us first go back to the interactions of the sterile part of the RH sneutrino.  Among these, only a four-point interaction or a LH sneutrino or a higgsino exchange contribute to RH sneutrino annihilation (see fig.(\ref{fig: interactions})).  They will be out of equilibrium as long as their interaction rate $\Gamma$ is smaller than the expansion rate $H=T^2/M_P$.  This necessitates for the 4-point interactions that
\beq
\Gamma_{4} \sim \lambda^4 T < H \Rightarrow T>10^{-33} \mathrm{GeV} 
\eeq
where we've taken $\lambda=10^{-13}$ (considering the observed neutrino mass splitting squared to be indicative of the neutrino masses).  This condition is obviously respected throughout the history of the Universe.  The LH sneutrino exchange and higgsino exchange have rates $\Gamma_{\tilde{\nu}_L}\sim \lambda^4a^4 T /m_{\tilde{\nu}_L}^4$ and $\Gamma_{\tilde{H}} \sim \lambda^4 T^3/m_{\tilde{H}}^2$; imposing that these be smaller than $H$ constrains the parameters such that
\bea
\Gamma_{\tilde{\nu}_L}&<&H \Rightarrow \frac{a^4}{m_{\tilde{\nu}_L}^4}<10^{33} T  \nn \\
\Gamma_{\tilde{H}}&<&H \Rightarrow\frac{T}{m_{\tilde{H}}^2}<10^{33} \mathrm{GeV}^{-1}  ~.
\eea
The first constraint is the strongest at low temperature while the second is strongest at high temperatures, so let us consider the first at the temperature no and the second at the highest reheating temperature that escapes the gravitino problem, $T_{RH} \sim 10^9$GeV \cite{gravitino_ellis, gravitino_khlopov, gravitino_ellis2}:
\bea \label{eq: constraints1}
a&<&10^{5}m_{\tilde{\nu}_L} \mathrm{GeV} \nn \\
m_{\tilde{H}}&>&10^{-12} \mathrm{GeV}~.
\eea
The second condition is certainly met; as for the first one, it is a quite mild condition on the trilinear coupling $a$ that can be easily met.  Hence these interactions do not generally reach equilibrium.  Moreover, before the electroweak phase transition, left-right mixing is out-of-equilibrium as well.  These could go through 4-points interactions, which we already know to be out of equilibrium; they could also go through higgs exchange, which has a rate of $\Gamma_{LR,h} \sim \lambda^2 a^2 g^2/T$ .  Imposing the rate to be smaller than $H$ all the way down to the electroweak transition, $T_{ew}$ forces
\beq
a<\sqrt{\frac{T_{ew}^3}{\lambda^2 M_P}} \sim 10^7 \mathrm{GeV}~,
\eeq
which again is easily respected.  Hence the RH sneutrinos never equilibrate with the plasma before the electroweak phase transition.

After the electroweak phase transition, we consider left-right mixing through mass insertion\footnote{Would left-right equilibration remain out of equilibrium after the electroweak phase transition, it would not allow for a quicker annihilation equilibrium and would only decrease the final relic density.  Hence we will consider that after the electroweak phase transition gauge and gaugino channels become open as it is a safe assumption that will also simplify the calculation.}.  The RH sneutrino now has an active part proportional to $(\lambda a v )/\left( m_{\tilde{\nu}_L}^2-m_{\tilde{\nu}_R}^2 \right)$.  There are now gauge and gaugino interactions that allow for RH sneutrino annihilations; their rates are respectively given by 
\bea
&\Gamma_{Z}& \sim  \frac{a^4 \lambda^4 v^4 g^2 T^5}{m_{Z}^4 \left( m_{\tilde{\nu}_L}^2-m_{\tilde{\nu}_R}^2 \right)^4} \nn \\
&\Gamma_{\tilde{W}}& \sim \frac{a^4 \lambda^4 v^4 g^2 T^3}{m_{\tilde{W}}^2 \left( m_{\tilde{\nu}_L}^2-m_{\tilde{\nu}_R}^4 \right)^2} ~.
\eea
Again, imposing these rates to be smaller than $H$ is turned into a constraint on the unknown parameters
\bea \label{eq: constraint2}
\frac{a}{\left( m_{\tilde{\nu}_L}^2-m_{\tilde{\nu}_R}^2 \right)}&<&10^6 \mathrm{GeV}^{-1} ~.
\eea
This constraint implies that a mass degeneracy between left- and right-handed sneutrinos might  eventually allow thermalisation of the RH sneutrino, depending on the size of the trilinear coupling.  For a trilinear coupling and a LH sneutrino mass of order $100$GeV, only a RH sneutrino mass greater than $\sim 100 \mathrm{GeV}-10 \mathrm{eV}$ (but smaller than $100$GeV for it to be the LSP) can spoil the non-equilibration; for the case of the mattergenesis mechanism of \cite{abelpage}, $a \sim 100$GeV and $m_{\sneu} \sim1$GeV, the LH sneutrino mass has to be between $1$ and $1,0001$ GeV to allow thermalisation.  The condition of eq.(\ref{eq: constraint2}) is thus fairly mild.

We can re-express eq.(\ref{eq: constraint2}) as
\bea
\frac{\lambda a v }{\left( m_{\tilde{\nu}_L}^2-m_{\tilde{\nu}_R}^2 \right)}<3\times 10^{-5}~.
\eea
This implies that the active part of the RH sneutrino (see eq.(\ref{eq: rm1})) has to be small in order not to allow thermalisation.  A LH part that's non-negligible when compared with the Yukawa coupling is still allowed however, and for this reason the gauge and gaugino couplings cannot be neglected altogether.

So it appears that as long as the active part of the RH sneutrino is kept under control, the (LSP) RH sneutrino never thermalises.  This is of little interest in a case where one is trying to simply have RH sneutrino DM with a relic density matching the observed one.  It is however of interest in the opposite case of DM created through mattergenesis; supposing that a certain mechanism has created $\Omega_{\sneu}$, we have shown up to now that this asymmetry will remain unaffected by the thermal plasma at later times.

\section{Dirac RH sneutrino relic density} \label{sec: relicdensity}
In the typical picture of a supersymmetric partner that decouples after becoming non-relativistic, the weak interactions of the RH sneutrino would generally produce a large relic density \cite{kamiongriest}.  However the fact that RH sneutrino annihilations never equilibrate steers us away from the conventional calculation, and as we'll see the final relic density can be in fact very low.  Here RH sneutrinos are created by the decays of the type $1 \to 2+\sneu$, where particles $1$ and $2$ are in equilibrium.  The relic density is given by solving the Boltzmann equation (see eg. \cite{kolbturner}), which states here that the time evolution of the number of RH sneutrinos $n_{\sneu}$ follows
\beq \label{eq: boltz}
\dot{n}_{\sneu}+3Hn_{\sneu}=\sum_{i}\mathrm{C_i}~, 
\eeq
where the term linear in $n_{\sneu}$ accounts for the expansion of the Universe, and $\sum_{i}\mathrm{C_i}$ is the 'collision' operator.  For the case at hand $\sum_{i}\mathrm{C_i}$ is given by
\beq \label{eq: cs}
C_i=\int \!\! \! \int \!\! \! \int \frac{d^3p_1}{\Pi_1} \frac{d^3p_2}{\Pi_2} \frac{d^3p}{\Pi_{\sneu}} \left( 2 \pi \right)^4 \delta^4\left(p_1-p_2-p_{\sneu}  \right) \left| M_i \right|^2 \left( 1\pm f_{\sneu}  \right)  \left(1 \pm f_2  \right) f_1 
\eeq 
where $\Pi_x = (2 \pi)^3 2 E_x$, $f_x$ are distribution functions (Fermi-Dirac or Bose-Einstein for particles $1$ and $2$), and there's a $C_i$ to be calculated for every decay channel using the corresponding $|M|^2$ as given on fig.(\ref{fig: interactions}).  Eq.(\ref{eq: boltz}) can be rephrased in terms more convenient for a numerical solution using the yield variable $Y \equiv n/s$, where $s$ is the entropy density of the Universe.  In terms of $Y$ the Boltzmann equation becomes
\bea \label{eq: boltzyield}
Y=\int_{T_{now}}^{T_{reh}}\frac{\sum_{i}C_i}{sHT}dT ~.
\eea
The yield variable is directly related to the density parameter through
\bea
\Omega_{relic}&=&\frac{\rho_{\sneu}/s}{\Omega_{crit}}=\frac{m_{\tilde{\nu_R}}Y}{3.5\times10^{-9}h^2} 
\eea
where $\Omega_{crit}$ and $h$ can be obtained using \cite{PDG}.  Going back to the collision operator, using the Maxwell-Boltzmann approximation  \cite{kolbturner} and integrating everything that can be integrated straightforwardly, we obtain
\bea
C_i = \frac{4}{\left( 2 \pi\right)^3}\int \!\!\! \int \!\!\! \int\frac{d\left(\cos\theta\right)d\left|\vec{p} \right| dE_1} {E_1 E_{\sneu} E_2\left| M_i \right|^2} \left|\vec{p}_{\sneu} \right|^2  \left|\vec{p_1} \right| f_1 \delta\left( E_1 - E_{\sneu} - E_2 \right) ~.
\eea
where $E_2$ now stands for $\sqrt{\left|\vec{p_1}-\vec{p}_{\sneu}\right|^2+m_2^2}$.  

We have solved eq.(\ref{eq: boltzyield}) for a number of sets of parameters that respect the constraints obtained in section \ref{sec: nontherm} (table \ref{table: diffmodels}).  The relic densities obtained are completely negligible compared to the observed DM density.  Using much larger trilinear couplings (but allowed by the non-thermalisation constraints) or degenerate sneutrino masses would  lead to large relic densities \cite{moroi}.  It is not the case, then, that large relic densities are simply impossible; they can indeed be achieved by a certain level of parameter tuning.  What we wish to suggest here, however, is that, as exemplified in table \ref{table: diffmodels}, for a large number of models that respect only mild constraints on the trilinear couplings and sneutrino masses, the relic density of RH sneutrino is indeed very small.  When this observation is coupled to the fact that the RH sneutrino is not thermal, it leaves wide open the possibility of DM generation through mattergenesis.

\begin{table} 
\begin{center}
\begin{tabular}{|c|c|c|c||c|}\hline 
$a$ & $m_{\tilde{\nu}_L}$&$m_{\tilde{\nu}_R}$ & $m_{\tilde{H}}$& $\Omega_{relic}$ \\
\hline
$100$ & $300$ & $100$ & $1000$ & $3 \times 10^{-5}$ \\
$1000$ & $300$& $100$ & $1000$ & $8 \times 10^{-3}$ \\
$2000$ & $300$& $1$ & $1000$ & $3 \times 10^{-3}$\\
$100$ & $300$ & $1$ & $1000$ & $3 \times 10^{-5}$\\
$100$ & $100$ & $1$ & $1000$ & $2\times 10^{-4}$\\
$100$ & $50$ & $1$ & $1000$ & $2 \times 10^{-3}$\\
\hline
\end{tabular}
\caption{Various set of parameters and the relic density they generate.  All masses are in GeV.  The fourth line corresponds to the mattergenesis model of \cite{abelpage}; the scenario suggested in \cite{dmkuzmin} only stated that a sneutrino mass of $O(1)$ GeV should be used (lines 3 to 7).  The scenario presented in \cite{mcdonald} requires a RH sneutrino mass of $\sim 100$GeV.  A large trilinear coupling or a close degeneracy of sneutrino masses enlarge the relic density.}   \label{table: diffmodels}
\end{center}
\end{table}

\begin{figure}[t]
\begin{center}
\includegraphics[width=0.4\textwidth]{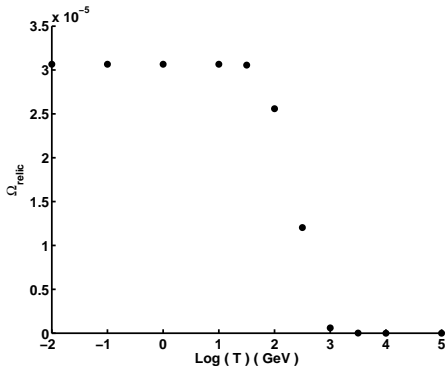}
\caption{Evolution of the RH sneutrino relic density as a function of temperature (time running backwards).  The parameters that have been used here are the ones in the fourth line of table \ref{table: diffmodels}.  The next-to-LSP will freeze-out at around typically $m_{NLSP}/20$, at which point the RH sneutrino relic density has already reached its final value.  Some time after the NLSP freeze-out the NLSP relic density will be 'dumped' into a RH sneutrino one, thus adding a (hopefully small - see text) 'step' to this plot.}
\label{fig: relicevolution}
\end{center}
\end{figure}

This is not the end of the story, however, because we have up to now only considered the case where all the other particles involved (apart from the RH sneutrino) are in equilibrium.  They will eventually freeze-out, however, and since none of them is the LSP, their relic density will eventually be 'dumped' into the RH sneutrino one.  In other words, the next-to-LSP (NLSP) (or MSSM-LSP) relic density should also be calculated and added to the RH sneutrino relic density \cite{moroi} (see fig.(\ref{fig: relicevolution})).  This is the one thing small Yukawas give no protection against: dumping of large amounts of RH sneutrino by the decay of a MSSM-LSP that would happen to have a relic density  comparable to the observed dark matter density.  Fortunately a low MSSM-LSP is still an open possibility.  Indeed it has been noticed recently that some LSP candidates usually considered to have 'correct' relic densities can have instead either much too high or - more importantly here - much too low relic densities (see for example \cite{arkani_relic} or \cite{belanger_relic} for an overview).  It should be mentioned, however, that no matter what their DM candidate is, any mattergenesis scenario has to assess the question of MSSM-LSP dark matter, because if indeed the MSSM with R-parity is a reality, then the LSP is necessarily a source of dark matter.

The left-right 'non-equilibration' before the electroweak phase transition was noticed in (non-supersymmetric) Dirac neutrinos in \cite{dicketal}, where it was argued that it can be used to construct a  leptogenesis mechanism in the neutrino sector.  The mechanism was used further in the MSSM in \cite{abelpage}, where this time it rendered possible the creation of both baryonic and dark matter through the Affleck-Dine mechanism, a possibility also noticed more recently in \cite{mcdonald}.  The parameters that yield a correct DM density in \cite{abelpage} ($a\sim100$GeV and $m_{\tilde{\nu}_R}\sim 1$GeV, no constraint on $m_{\tilde{\nu}_L}$) are 'generic' enough to respect all the constraints of section \ref{sec: nontherm} and produce a small relic density (see table \ref{table: diffmodels}).  This ensures that the RH sneutrino density created around reheating time by the Affleck-Dine mechanism remains as such and can be straightforwardly taken as the DM density, as was actually assumed in \cite{abelpage}.  It is interesting to note that while \cite{abelpage,mcdonald} showed that the smallness of the Yukawa coupling provides a simple way of creating a DM density, what we've shown here is that it is also a built-in protection of this density.  We also mention in table \ref{table: diffmodels} the scenario of \cite{dmkuzmin}, in which a new, heavy Majorana fermion $X$ decays to a RH sneutrino and a SM fermion.

\section{Conclusion}
The Dirac RH sneutrino LSP is a natural candidate for DM within a mattergenesis scenario.  Unless the trilinear coupling $a$ or the degeneracy between left- and right-handed sneutrino become very large, the RH sneutrino never thermalises throughout the history of the Universe.  This is a desirable characteristic in most mattergenesis scenarios as it ensures that a DM energy density created in the early Universe is not reprocessed at later times by fast annihilations.  Moreover the relic density of RH sneutrino LSP is generally much smaller than the observed DM density, which implies that no large amount of additional DM is added to a density created earlier.  Again this remains true as long as both trilinear coupling and sneutrino mass degeneracy are kept under reasonable control.  In the mattergenesis scenarios suggested in previous works \cite{abelpage,mcdonald, dmkuzmin}, the constraints obtained here are easily respected.  As such the MSSM+$\sneu$ appears as the minimal extension to the MSSM that allows for DM to have mattergenesis as its source.

\section{Acknowledgments}
The work of V.P. is funded by the Fonds Nature et Technologies du Qu\'ebec.  V.P. would like to thank Steven Abel for helpful discussions and comments throughout this work, and is grateful to Jean-Marie Fr\`ere and Michel Tytgat for their hospitality at Universit\'e Libre de Bruxelles where the first part of this work was carried.

\bibliographystyle{unsrt}
\bibliography{paper3}

\end{document}